\documentclass[conference]{IEEEtran}
\IEEEoverridecommandlockouts
\usepackage{cite}
\usepackage{lscape}
\usepackage{pdfpages}
\usepackage[figuresleft]{rotating}
\usepackage{amsmath,amssymb,amsfonts}
\usepackage{algorithmic}
\usepackage{graphicx}
\usepackage{textcomp}
\usepackage{xcolor}
\usepackage{soul}
\usepackage{multirow}
\usepackage{booktabs}
\usepackage{adjustbox}
\usepackage{amsmath}
\usepackage{color, colortbl}
\usepackage{tikz}
\usepackage{hyperref}
\def\BibTeX{{\rm B\kern-.05em{\sc i\kern-.025em b}\kern-.08em
    T\kern-.1667em\lower.7ex\hbox{E}\kern-.125emX}}

\newcommand\copyrighttext{%
    \centering
  \tiny
  \copyright~This paper is under consideration at Pattern Recognition Letter}
\newcommand\copyrightnotice{%
\begin{tikzpicture}[remember picture,overlay]
\node[anchor=south,yshift=10pt] at (current page.south) {\fbox{\parbox{\dimexpr\textwidth-\fboxsep-\fboxrule\relax}{\copyrighttext}}};
\end{tikzpicture}%
}    

\begin{document}

\newcommand{\highl}[1]{{#1}}
\definecolor{Gray}{gray}{0.9}

\title{An Open-Set Recognition and Few-Shot Learning Dataset for Audio Event Classification in Domestic Environments}

\author{\IEEEauthorblockN{Javier Naranjo-Alcazar}
\IEEEauthorblockA{
\textit{Visualfy, Universitat de Valencia}\\
Benisano, Spain \\
javier.naranjo@visualfy.com}
\and
\IEEEauthorblockN{Sergi Perez-Castanos}
\IEEEauthorblockA{
\textit{Visualfy}\\
Benisano, Spain \\
sergi.perez@visualfy.com}
\and
\IEEEauthorblockN{Pedro Zuccarello}
\IEEEauthorblockA{
\textit{Visualfy}\\
Benisano, Spain \\
pedro.zuccarello@visualfy.com}
\and
\IEEEauthorblockN{Ana M. Torres}
\IEEEauthorblockA{
\textit{Universidad de Castilla-La Mancha}\\
Cuenca, Spain \\
ana.torres@uclm.es}
\and
\IEEEauthorblockN{Jose J. Lopez}
\IEEEauthorblockA{
\textit{Universitat Polit\`ecnica de Val\`encia}\\
Valencia, Spain \\
jjlopez@dcom.upv.es}
\and
\IEEEauthorblockN{Francesc J. Ferri}
\IEEEauthorblockA{
\textit{Universitat de Valencia}\\
Burjassot, Spain \\
francesc.ferri@uv.es}
\and
\IEEEauthorblockN{Maximo Cobos}
\IEEEauthorblockA{
\textit{Universitat de Valencia}\\
Burjassot, Spain \\
maximo.cobos@uv.es}
}

\maketitle

\begin{abstract}
The problem of training with a small set of positive samples is known as few-shot learning (FSL). It is widely known that traditional deep learning (DL) algorithms usually show very good performance when trained with large datasets. However, in many applications, it is not possible to obtain such a high number of samples. In the image domain, typical FSL applications include those related to face recognition. In the audio domain, music fraud or speaker recognition can be clearly benefited from FSL methods. This paper deals with the application of FSL to the detection of specific and intentional acoustic events given by different types of sound alarms, such as door bells or fire alarms, using a limited number of samples. These sounds typically occur in domestic environments where many events corresponding to a wide variety of sound classes take place. Therefore, the detection of such alarms in a practical scenario can be considered an open-set recognition (OSR) problem. To address the lack of a dedicated public dataset for audio FSL, researchers usually make modifications on other available datasets. This paper is aimed at providing the audio recognition community with a carefully annotated dataset\footnote{https://zenodo.org/record/3689288} for FSL in an OSR context comprised of 1360 clips from 34 classes divided into \emph{pattern sounds} and \emph{unwanted sounds}. To facilitate and promote research on this area, \highl{results with state-of-the-art baseline systems} based on transfer learning are also presented. 
\end{abstract}


\copyrightnotice

\section{Introduction}\label{sec:intro}

The automatic classification of audio clips is a research area that has grown significantly in the last few years \cite{b4, b5, b6, b7, b8}. The research interest in these algorithms is motivated by their numerous applications, such as audio-based surveillance, hearing aids, home assistants or ambient assisted living, among others.
\highl{In this context, smart acoustic monitoring systems can be used to assist elderly and disabled people~\cite{polap2018human, Cobos2016}, or to enable safer user interactions based on audio processing \cite{polap2016neuro}.}
While deep learning-based methods have shown outstanding results in many areas, especially in traditional ones such as image classification/segmentation \cite{b15, b16} or speech recognition, these remarkable results are based on data-intensive strategies and algorithms. In contrast, few-shot learning (FSL) tackles the problem of learning with few samples per class. FSL approaches gained focus when trying to address intra-class classification in the face recognition context \cite{b17}, including applications such as access control and identity verification \cite{b22,b23,b24}. In order to tackle this problem, loss functions such as ring loss \cite{b38} or center loss \cite{b39} have been proposed, together with different based embedding network architectures such as siamese \cite{b18, b19} and triplet \cite{b20, b21}. These loss functions are aimed at solving convergence issues both in siamese and triplet networks, which also require a careful training procedure to choose appropriately the pairs or triplets used. Solutions of this kind have been tested in the image domain with datasets such as MegaFace, containing around one million samples \cite{kemelmacher2016megaface}.

\begin{figure*}[ht!]
\centering
\includegraphics[scale=0.60]{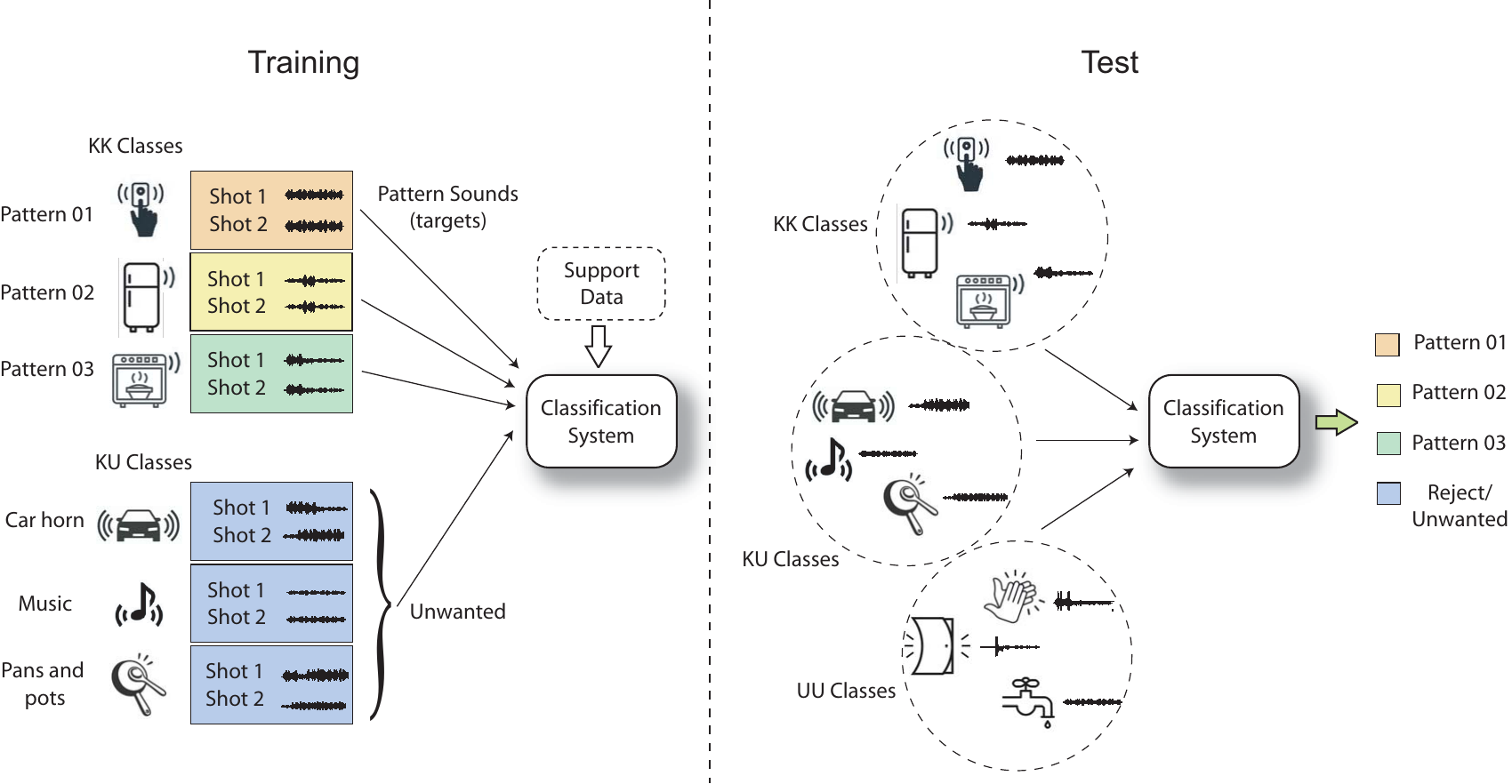}
\caption{Training and test under FSL and OSR conditions. In the training stage (left), only a few examples (shots) are available for each class, where some classes are targets to be recognized (KK classes) and others are unwanted classes to be rejected (KU classes). In the test stage (right) the system receives as input examples from the target classes and also from unwanted ones, where new classes different from the ones seen during training (UU classes) are also present.}
\label{fig:scheme}
\end{figure*}

\highl{Another practical issue arising in many real-world intelligent audio applications is open-set recognition (OSR)~\cite{b25}. This problem occurs when a system has to face unfamiliar situations for which it has not been trained. A system prepared for OSR should be capable of correctly classifying examples corresponding to classes seen during the training stage while rejecting examples corresponding to new, previously unseen classes. OSR has been addressed in the past by applying modifications to classical machine learning algorithms such as support vector machines \cite{scheirer2012toward, scheirer2014probability} or nearest neighbour classification \cite{junior2017nearest}. In the last years, deep learning solutions for OSR have also started to emerge, such as OpenMax~\cite{bendale2016towards}, deep open classifier (DOC) \cite{shu2017doc} or competitive overcomplete output layer (COOL) \cite{kardan2017mitigating}. However, despite its practical interest, this particular type of recognition problem is not so popular, evidencing the need for further research in this direction \cite{b26, b27}.}

\highl{The problems of FSL and OSR appear frequently in smart acoustic applications. For example, a given user may be exposed to several alerts or beeps at his home, emitted by different domestic appliances (e.g. oven and refrigerator). A smart system to differentiate between both alerts should not classify both sounds into a single ``alarm" class, but should be capable of identifying correctly those specific \emph{pattern sounds}. However, only a limited number of examples recorded by the user may be available for training. In addition, the system should neglect or discard the variety of possible sounds appearing in a domestic environment. Therefore, there is a need to design machine learning systems trained with a small number of audio examples capable of both identifying the classes of interest  (FSL) while rejecting the sounds coming from other unexpected sources (OSR).}

\highl{A diagram of the conditions under which training and testing are performed within a FSL+OSR context is shown in Fig.~\ref{fig:scheme}. The FSL condition is reflected by the small number of examples (shots) available during the training stage. On the other hand, the OSR condition is accounted by letting the system learn from examples corresponding to unwanted (non-target) classes. Since the number of examples is clearly insufficient, usually some meta-learning strategy and support data is needed to let the system learn to discriminate among data and exploit better the information provided by the available shots. In the test stage, the system is confronted towards examples pertaining either to target classes or to unwanted ones. Such unwanted examples might belong to the group of unwanted classes seen during the training stage, but they may also belong to new unseen classes, which makes the problem even more challenging. The classification system should be capable of identifying the target classes and to reject the unwanted ones. Following the OSR nomenclature (cf. Sec.~\ref{sec:exp_setup}), the involved groups of classes are denoted as KK, KU and UU in Fig.~\ref{fig:scheme}.}

\begin{figure*}[ht!]
\centering
\includegraphics[scale=0.30]{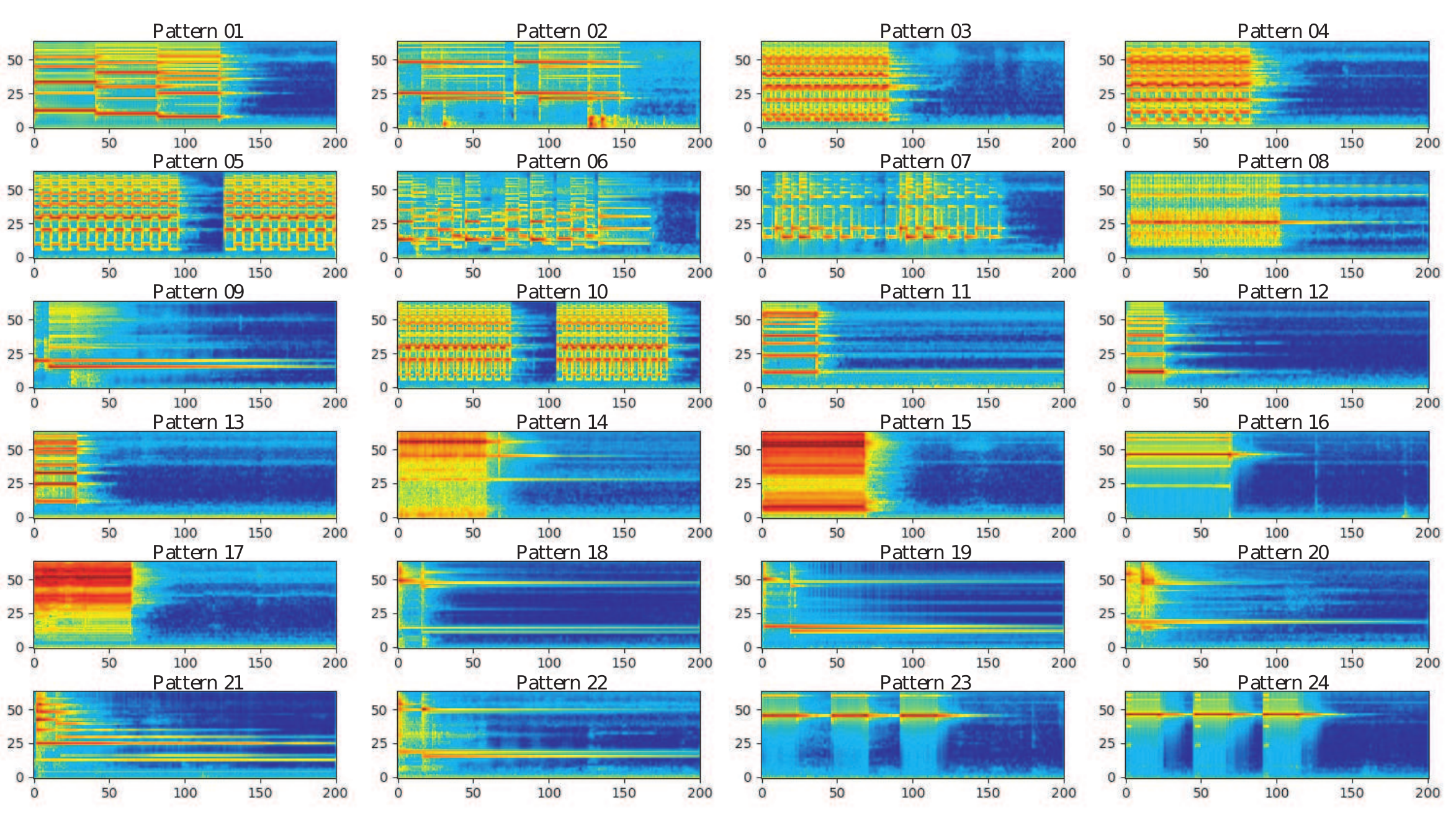}
\caption{Log-Mel spectrogram of the sounds in \emph{pattern sounds} category. One sample sound per class is shown. The horizontal axis corresponds to the time frame and the vertical axis to the Mel frequency band.}
\label{fig:pattern_specs}
\end{figure*}

\par Standard public datasets are a very important aspect of automatic classification research, whether it is based on deep, few-shot or any other learning technique. Several audio datasets have been released in order to validate different types of algorithms \cite{piczak2015esc, salamon2014dataset, gemmeke2017audio, Mesaros2018_DCASE}. To address the lack of a dedicated public dataset for audio FSL, researchers usually make modifications on other available datasets \cite{b3, continual_few_shot}. Thus, the dataset presented in this paper is aimed at facilitating research on FSL for audio event classification. A domestic environment is considered, where a particular sound must be identified from a set of \emph{pattern sounds}, all belonging to a general "audio alarm" class. The challenge resides in detecting the target pattern by using only a reduced number of examples. Moreover, a relevant aspect of the present dataset is that it allows to evaluate algorithms intended for open-set recognition (OSR). When running inference in a real-world application, the classification system will have to predict on samples from classes unseen during the training stage. To account for openness conditions, the dataset provides as well a folder of \emph{unwanted sounds} containing audio samples from different subclasses which are not considered to be audio alarms or pattern sounds. An optimal FSL+OSR system would be able to correctly identify all the instances belonging to the different pattern sounds by using only a few training examples, while rejecting all the examples pertaining to the general unwanted class. A preliminary version of this dataset has already been used in a previous work \cite{naranjo2020open}.

\highl{One of the main motivations} of this paper is to facilitate open research in the field of audio-oriented FSL and OSR. For this purpose, the dataset is accompanied by \highl{two baseline systems based on transfer learning. One of them uses embeddings from the well known L$^3$-net network \cite{b30} and the other uses the YAMNet network, both pre-trained on Audioset\cite{gemmeke2017audio} and released by Tensorflow\footnote{https://github.com/tensorflow/models/tree/master/research/audioset/yamnet}}. The concept of \emph{openness}, as defined by Scheirer \emph{et al.} in \cite{Scheirer_2014_TPAMIb}, will be used to evaluate the performance of such baseline systems under different openness conditions and a different number of training examples.

The paper is structured as follows. Section \ref{sec:dataset} presents the general structure and organization of the dataset. Section \ref{sec:exp_setup} describes the experimental set-up used for evaluating the baseline systems that accompany the dataset, which are presented in detail in Section \ref{sec:baseline}. The results of the experiments are discussed in Section \ref{sec:results}. Finally, the conclusions of this work are summarized in Section \ref{sec:conclusion}.

\section{Dataset}\label{sec:dataset}

\par The dataset is divided into 34 taxonomic classes. These 34 classes are classified into one of two main sub-categories: \textit{pattern sounds} and \textit{unwanted}. \highl{The dataset is completely balanced, as every class contains exactly the same number of audio examples.}

\begin{itemize}
    \item \textit{Pattern sounds category}: it is comprised of a total of 24 classes, each one being a different type of audio alarm (e.g. fire alarms or door bells). Each \textit{pattern sound} class has 40 audio clips.
    \item \textit{Unwanted category}: it is comprised of a total of 10 different classes, each one representing everyday domestic audio sources: \textit{car horn, clapping, cough, door slam, engine, keyboard tapping, music, pots and pans, steps and water falling}. Each of these \textit{unwanted} classes has 40 audio clips.
\end{itemize}

Moreover, a $k$-fold configuration is provided in order to check the generalisation of the results. The number of folds ($k$) for cross-validation depends on the number of shots used for learning. That means, when training with 4 shots, the number of folds is $k=10$. For 2 shots, $k=20$. Consequently, there are 40 folds for 1 shot.

All the audio sequences have a duration of 4 seconds and have been recorded using a single audio channel with a sample rate of 16~kHz and 16~bits~per~sample with \highl{pulse code modulation} (PCM) codification. The dataset along with other detailed information is available in the following link\footnote{https://zenodo.org/record/3689288 - Full unrestricted and anonymous access has been set during the revision process of this manuscript. Report any access problem to journal staff.}. For illustrative purposes, the log-Mel spectrograms corresponding to audio examples belonging to the different classes of the \emph{pattern sounds} category are represented in Fig.~\ref{fig:pattern_specs}. The log-Mel spectrogram has been calculated with a window size of $40$ ms, an overlap of $50\%$ and $64$ frequency bands. All frequency bins have been normalized with zero mean and standard deviation equal to one.

\begin{figure*}[ht!]
\centering
\includegraphics[scale=0.30]{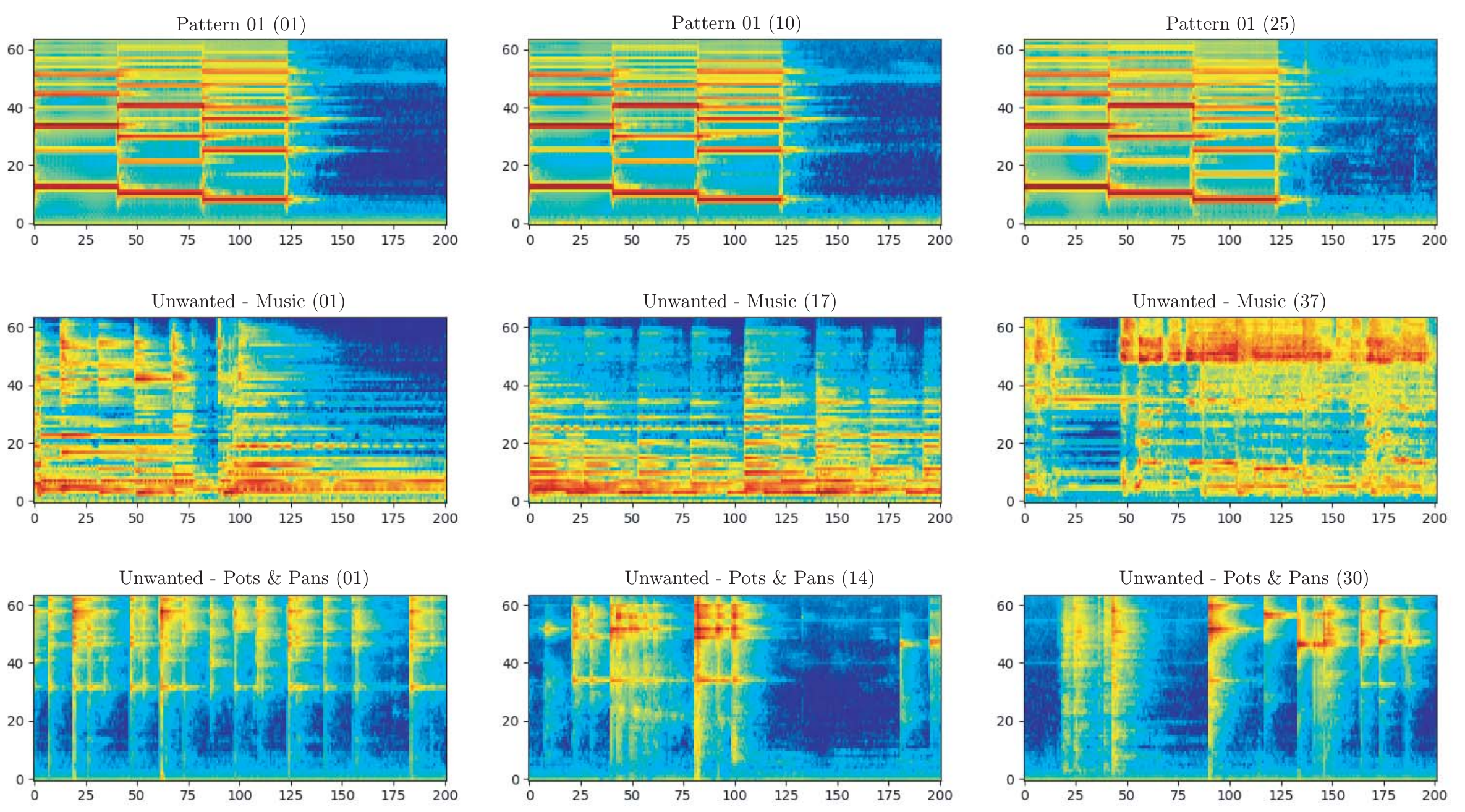}
\caption{Comparison of log-Mel spectrograms from within-class examples corresponding to a pattern sound class (first row) and two classes from the unwanted category (second and third rows). Note that the inter-class variability of the examples in the first row is considerably smaller than in the rest of examples extracted from more general sound classes. The number in parenthesis denotes the example index within the class.}
\label{fig:examples_unwanted}
\end{figure*}

Examples corresponding to the same pattern sound class are expected to share similar characteristics, while those from unwanted classes tend to show higher variability, as they come from more general sound events.
This can be observed in Fig.~\ref{fig:examples_unwanted}, which shows three examples from three classes: one pattern sound class and two unwanted classes.

\section{Experimental setup}\label{sec:exp_setup}

\par The aim of the experiments is to test the performance of the baseline system over the proposed dataset considering both OSR and FSL conditions. The evaluation under open-set conditions is based on the concept of \emph{openness} \cite{Scheirer_2014_TPAMIb}. For this purpose, the \emph{pattern sounds} and \emph{unwanted} categories detailed in Sect.~\ref{sec:dataset} are further subdivided as follows:
\begin{itemize}
    \item \textit{Known Known} (KK) classes: KK are the classes whose audios have been used for training/validation labeled as positive events to be recognized by the system. In the context of this work, KK classes would match the \emph{pattern sounds} category.
    \item \textit{Known Unknown} (KU) classes: KU are the classes whose audios have been used for training/validation, but labeled as unwanted categories so that they are not classified as positive events during testing. In this work, KU classes would be represented by a subset of the \emph{unwanted} classes.
    \item \textit{Unknown Unknown} (UU) classes: as in the case of KU classes, UU classes are a subset of the \emph{unwanted} group. The difference between KU and UU is that the audios in UU classes are not used for training/validation; instead, they are only used in the testing phase. It is expected that audios in UU subset will be classified as unwanted by the system after the training/validation stage has been finished.
\end{itemize}

The openness, $O^*$, can be calculated using the following formula \cite{geng2018recent}:
\begin{equation}
O^* = 1-\sqrt{\frac{2 \times |C_{TR}|}{|C_{TR}| + |C_{TE}|}},
\label{for:openness}
\end{equation}
where $C_{TR}$ is the set of classes used during training, $C_{TR} = \textrm{KK} \cup \textrm{KU}$, and $C_{TE}$ corresponds to the set of classes used in testing phase, $C_{TE} = C_{TR} \cup \textrm{UU}$. Openness values are bounded to the range $0 \leq O^* < 1$. When $C_{TR}=C_{TE}$, $O^*$ reaches its minimum value ($O^*=0$), meaning that, during testing, the algorithm is not required to face events that belong to classes unseen during training. On the contrary, as the difference between $|C_{TE}|$ and $|C_{TR}|$ becomes larger, with $|C_{TE}|>|C_{TR}|$, the openness tends to approach to its maximum value: $O^* \xrightarrow{} 1$. This means that, during testing, the system needs to reject events belonging to classes unseen during training.

\begin{table}[]
\centering
\caption{Number of classes of each configuration and the corresponding openness value. }
\begin{tabular}{ccccccc}
\toprule
\scriptsize
{\bf Pattern Sounds} & 
{$|KK|$} & {$|KU|$} & {$|UU|$} 
& {$|C_{TR}|$}  & {$|C_{TE}|$}  &{ $O^*$} \\

\midrule
\multirow{3}{*}{Full set} & \multirow{3}{*}{24}         & 10         & 0          & 34         & 34         & 0     \\ 
    &  & 5          & 5          & 29         & 34         & 0.04  \\ 
      &  & 0          & 10         & 24         & 34         & 0.09  \\ \midrule
\multirow{3}{*}{Trios} & \multirow{3}{*}{3}          & 10         & 0          & 13         & 13         & 0     \\
        &  & 5          & 5          & 8          & 13         & 0.13  \\ 
    &      & 0          & 10         & 3          & 13         & 0.39 \\ \bottomrule
\end{tabular}
\label{tab:classes_openness}
\end{table}

In a first batch of experiments, all 24 \emph{pattern sounds} classes have been used together as KK classes. In a second batch, \emph{pattern sounds} have been selected in 8 groups of 3 classes each (8 trios, as later identified in Section \ref{sec:results}), therefore, only 3 classes per run have been used as KK. 
The particular classes in each trio have been selected to cover different everyday situations ranging from very different sounds as (1,9,17) to more similar ones as (4,5,16). This second batch reflects a more realistic scenario where the number of classes in the union of KU and UU subsets (KU $\cup$ UU) outnumbers the classes in the KK group. Besides, the experimental setup was designed to have several degrees of freedom taking into account the number of positive audio samples used for training (also called shots) and different values of openness. Experiments with one, two and four shots have been carried out. In order to obtain different values of openness, the ratio given by the number of KU classes and the number of UU classes has been set to 10/0, 5/5 and 0/10. This results in $O^* \in \{0,\, 0.04,\, 0.09\}$ for the first batch of experiments and $O^* \in \{0, \, 0.13, 0.39\}$ for the second batch. Table \ref{tab:classes_openness} summarizes the details related to the two types of experiments described above. \highl{Note that, in all cases, we have a completely balanced classification problem with $|KK|$ classes, with a reject option.}

\section{Baseline systems}\label{sec:baseline}

\par Transfer learning \cite{b6, b17} is a well-known technique that takes advantage of prior knowledge from previously trained neural networks to tackle new problems. New training samples are mapped into a specific domain that has been calculated with some other data. With such an approach, a pre-trained neural network is used as a feature extractor, where the activations extracted from its inner layers are used to face new classification problems. Such internal representations are commonly known as embeddings or deep features \cite{Martin2018ICSP, Martin2019ICA}, and are considered as a powerful alternative to typical hand-crafted features such as Mel-Frequency Cepstral Coefficients (MFCCs) \cite{Martin2016}. This approach shows excellent results in contexts where, as in the case of FSL, a small training dataset is available \cite{b17}. \highl{In this work, two systems based on pre-trained networks have been used for comparison purposes. The details of the implementations are described in the following subsections.}

\subsection{L$^3$-net}

\par The first system is based on a pre-trained network used for feature extraction, as provided by L$^3$-net\footnote{https://openL$^3$-net.readthedocs.io/en/latest/tutorial.html} embeddings \cite{b30}. L$^3$-net is a neural network trained with two specific partitions of Audioset from subsets corresponding to environmental and music videos. The parameters of the embedding were set as follows:
\begin{itemize}
    \item \textit{content\_type} = "music"
    \item \textit{input\_repr} = "mel256"
    \item \textit{embedding\_size} = 512
    \item \textit{hop\_size} = 0.5
\end{itemize}

\par For the computation of the L$^3$-net embeddings, each audio clip is divided into 1-second segments with a hop size of 0.5 seconds. Taking into account the 1 second analysis window used by L$^3$-net, the above parameters lead to an embedding matrix of size $512\times7$\cite{b30}. We summarize this output by averaging across the temporal dimension, resulting in a $512\times1$ column-vector representation. Details on its architecture are given in Table~\ref{tab:l3_architecture}.

\begin{figure*}[ht!]
\centering
\includegraphics[scale=0.22]{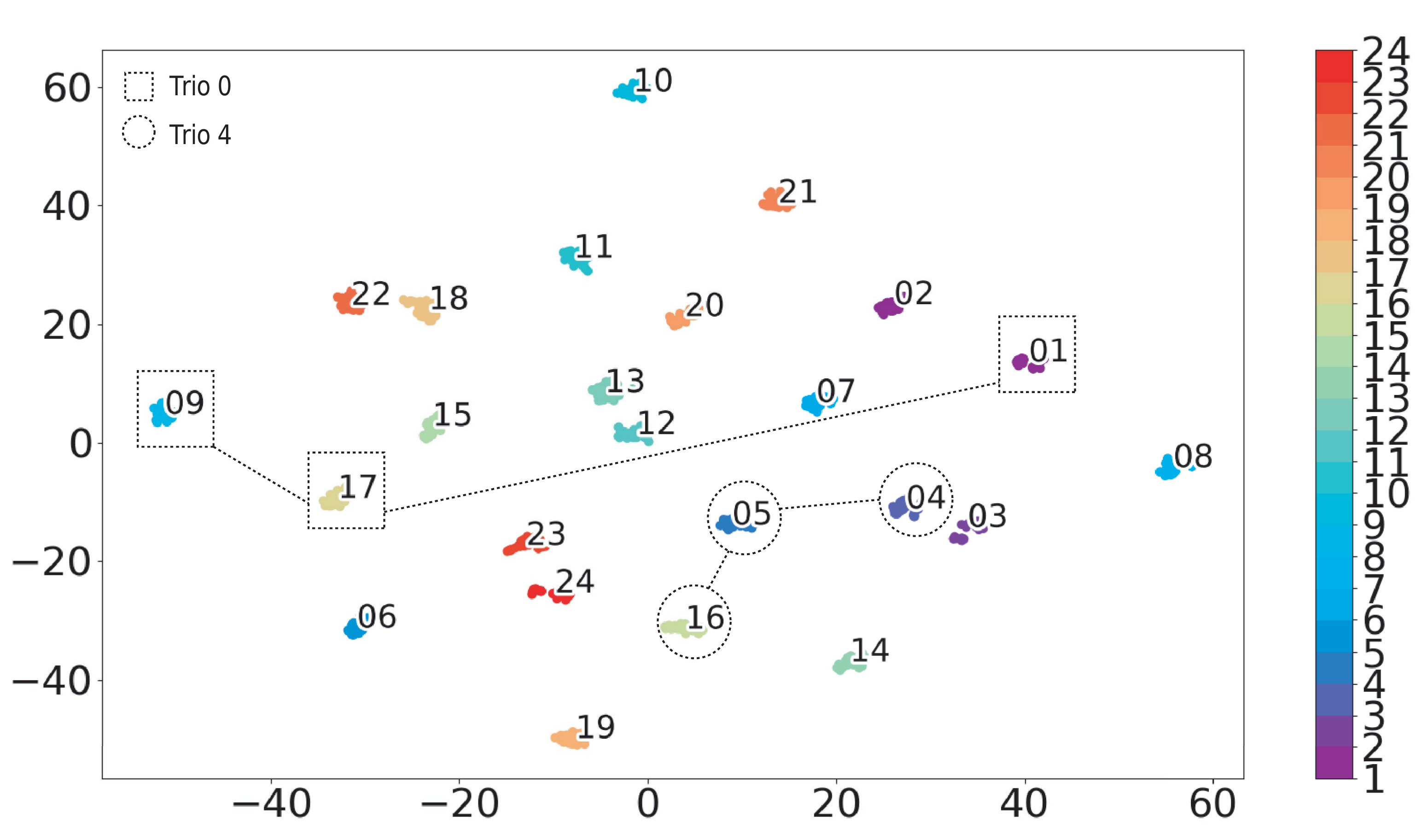}
\caption{t-SNE mapping from L$^3$-net representation of 24 KK categories.}
\label{fig:pattern_tsne}
\end{figure*}

For visualization purposes, a t-SNE mapping of the computed feature representations for the KK classes is shown in Fig. \ref{fig:pattern_tsne}. This representation gives insight about the mapping of the different classes into the feature space. Note that the t-SNE representation captures faithfully the similarity existing among examples of the same pattern sound class, leading to visibly condensed clusters. Moreover, pattern sound classes having a similar spectro-temporal structure can be identified to be closely separated in the t-SNE mapping. For example, classes 3-4 or 23-24 appear close to each other and, according to Fig.~\ref{fig:pattern_specs}, there is an obvious similarity between them.

\subsection{YAMN\lowercase{et}}\label{subsec:yamnet}

\highl{The implementation of this system follows the same procedure as the one using L$^3$-net embeddings.} 
\highl{In this case, the audio pre-processing is based on log-Mel spectrograms using 64 frequency bands and a frame size of 0.96 s with 50\% overlap. For 4 second audio clips, the extracted audio embeddings have a shape of $1024 \times 8$. As with L$^3$-net, the mean across the temporal axis is computed to flatten such output. YAMNet has also been trained using Audioset \cite{gemmeke2017audio}. Its architecture is summarized in Table~\ref{tab:yamnet_architecture}.}

\begin{table}[]
\centering
\caption{Full framework \highl{using L$^3$-net as feature extractor} divided per layers. Convolutional layers are indicated using $\#$ to represent the number of filters and the values in brackets as the kernel sizes. \highl{Conv2D defines a convolutional block including batch normalization (BN) and ReLU activation.} This architecture is explained with more detail in \cite{arandjelovic2017look}. These layers are frozen as are only used as feature extractor. The layers in bold correspond to the additions to the L$^3$-net architecture. \highl{All dense layers have ReLU activation except the last one, which is set to sigmoid.}
The number of output units also vary depending on the number of KK classes to be targeted.}
\begin{tabular}{c}
\rowcolor{Gray}\emph{spectrogram input}\\
\midrule
Conv2D($\#64$, $(3,3)$) $\times$ 2  \\
\midrule
MaxPooling $(2,2)$                           \\
\midrule
Conv2D($\#128$, $(3,3)$) $\times$ 2 \\
\midrule
MaxPooling $(2,2)$                         \\
\midrule
Conv2D($\#256$, $(3,3)$) $\times$ 2 \\
\midrule
MaxPooling $(2,2)$                         \\
\midrule
Conv2D($\#512$, $(3,3)$) $\times$ 2 \\
\midrule
MaxPooling $(32,24)$                        \\
\midrule
\rowcolor{Gray}\emph{L$^3$-net embedding}\\
\midrule
\textbf{GlobalAveragePooling()}                   \\
\midrule
\rowcolor{Gray}\emph{feature}    \\
\midrule
\textbf{Dense($512$)}                               \\
\midrule
\textbf{Dense($128$)}                               \\
\midrule
\textbf{Dense($3/24$, activation=sigmoid)}     \\
\midrule
\rowcolor{Gray}\emph{KK probabilities}
\end{tabular}
\label{tab:l3_architecture}
\end{table}

\begin{table}[]
\centering
\caption{\highl{Summary of the YAMNet architecture, using the same notation as in Table~\ref{tab:l3_architecture}. New blocks Separable-Conv2D are formed by depthwise and pointwise convolutional layers, each followed by BN and ReLU, as originally proposed in \cite{howard2017mobilenets}.}}
\begin{tabular}{c}
\rowcolor{Gray}\emph{spectrogram input}\\
\midrule
Conv2D($\#32$, $(3,3)$)  \\
\midrule
Separable-Conv2D($\#64$, $(3,3)$)                           \\
\midrule
Separable-Conv2D($\#128$, $(3,3)$) $\times$ 2 \\
\midrule
Separable-Conv2D($\#256$, $(3,3)$) $\times$ 2                   \\
\midrule
Separable-Conv2D($\#512$, $(3,3)$)  $\times$ 6    \\
\midrule
Separable-Conv2D($\#1024$, $(3,3)$) $\times$ 2     \\
\midrule
GlobalAveragePooling() \\
\midrule
\rowcolor{Gray}\emph{YAMNet embedding}\\
\midrule
\textbf{GlobalAveragePooling()}                   \\
\midrule
\rowcolor{Gray}\emph{feature}    \\
\midrule
\textbf{Dense($512$)}                               \\
\midrule
\textbf{Dense($128$)}                               \\
\midrule
\textbf{Dense($3/24$, activation=sigmoid)}     \\
\midrule
\rowcolor{Gray}\emph{KK probabilities}
\end{tabular}
\label{tab:yamnet_architecture}
\end{table}

\subsection{System classifier}

\par For the classification task, a multi-layer perceptron with two fully-connected hidden layers with 512 and 128 units respectively was implemented as in \cite{b30}. \highl{This neural network is fed with either YAMNet or L$^3$-net embeddings independently}. All activation units are ReLUs. The output layer has 24 or 3 units (each one corresponding to a class of \textit{pattern sounds}) with sigmoid activation function. Labels that correspond to different \textit{unwanted sound} audio clip subcategories are set to a 24/3-zero vector. This indicates the absence of any \textit{pattern sounds} category.
Adam optimizer \cite{b31} was used. The loss function during training was binary cross-entropy and the evaluation metric was categorical accuracy. \highl{At test time, an audio clip is classified as known, or \emph{pattern sound}, when the corresponding output probability ranks the highest and above a threshold with value $0.5$. In the case where this threshold is exceeded by more than one class, the system predicts the class having the highest detection probability.} 
Results for this baseline framework \highl{using the two pre-trained networks} are discussed in Section \ref{sec:results}. \highl{The code for replicating the results is fully available\footnote{\url{https://github.com/Machine-Listeners-Valencia/fsl_osr_dataset_baseline}}.}

\begin{table*}[]
\centering
\small
\caption{Baseline system average accuracies (\%) and corresponding standard deviations (not shown for $ACC_{w}$) with 24 KK classes using L$^3$-net network. Shots indicates the number of training examples per class. 
}
\setlength\tabcolsep{1pt}
\begin{tabular}{ccc >{}cccc>{}ccc>{}c}
\toprule
&   & \multicolumn{9}{c}{Openness coefficient}  \\ \cmidrule(lr){2-11}
Shots & \multicolumn{3}{c}{$O^* = 0$} & \multicolumn{4}{c}{$O^* = 0.04$} & \multicolumn{3}{c}{$O^* = 0.09$} \\ \cmidrule(lr){1-11}
\multicolumn{1}{l}{} &  $ACC_{KK}$  & $ACC_{KU}$  & $ACC_w$ & $ACC_{KK}$  & $ACC_{KUU}$ & $ACC_{UU}$ & $ACC_w$ & $ACC_{KK}$    & $ACC_{UU}$   & $ACC_w$ \\ \cmidrule(lr){2-4} \cmidrule(lr){5-8} \cmidrule(lr){9-11}
1   & 13.8$\pm$12.9   &  99.8$\pm$1.0  & 56.8 & 57.7$\pm$8.4  & 90.4$\pm$5.4 & 84.8$\pm$9.8   & 74.1  & 60.1$\pm$7.8  & 39.6$\pm$13.4  & 49.9     \\ \cmidrule(lr){2-11}
 2    &  81.1$\pm$5.5   & 99.4$\pm$0.8 & \textbf{90.3} & 83.2$\pm$4.8  & 90.2$\pm$5.1 & 82.5$\pm$9.6  & \textbf{86.7}  & 83.3$\pm$5.6 & 33.3$\pm$11.6 & 58.3 \\ \cmidrule(lr){2-11}
 4    &  94.8$\pm$2.2   & 99.6$\pm$0.4  & \textbf{97.2}  &    94.3$\pm$2.2   & 88.3$\pm$5.7 & 79.4$\pm$9.5  & 91.3  & 94.8$\pm$2.4     & 26.1$\pm$10.1       & 60.5  \\ \bottomrule
\end{tabular}
\label{tab:results_op}
\end{table*}

\begin{table*}[]
\centering
\small
\caption{Baseline system average accuracies (\%) and corresponding standard deviations (not shown for $ACC_{w}$) with 24 KK classes using YAMNet network. Shots indicates the number of training examples per class. 
}
\setlength\tabcolsep{1pt}
\begin{tabular}{ccc >{}cccc>{}ccc>{}c}
\toprule
&   & \multicolumn{9}{c}{Openness coefficient}  \\ \cmidrule(lr){2-11}
Shots & \multicolumn{3}{c}{$O^* = 0$} & \multicolumn{4}{c}{$O^* = 0.04$} & \multicolumn{3}{c}{$O^* = 0.09$} \\ \cmidrule(lr){1-11}
\multicolumn{1}{l}{} &  $ACC_{KK}$  & $ACC_{KU}$  & $ACC_w$ & $ACC_{KK}$  & $ACC_{KUU}$ & $ACC_{UU}$ & $ACC_w$ & $ACC_{KK}$    & $ACC_{UU}$   & $ACC_w$ \\ \cmidrule(lr){2-4} \cmidrule(lr){5-8} \cmidrule(lr){9-11}
1   & 64.4$\pm$3.7   &  95.8$\pm$2.6  & \textbf{80.1} & 65.6$\pm$3.3  & 91.0$\pm$4.2 &  89.4$\pm$5.7   & \textbf{78.3}  & 66.9$\pm$3.2  & 47.3$\pm$13.1  &   \textbf{57.1}   \\ \cmidrule(lr){2-11}
 2    &  78.8$\pm$2.3   & 97.6$\pm$1.9 & 88.2 & 79.3$\pm$2.3  & 91.8$\pm$4.2 & 87.6$\pm$6.2  & 85.6  & 80.4$\pm$2.3 & 41.7$\pm$11.7 & \textbf{61.1} \\ \cmidrule(lr){2-11}
 4    &  90.8$\pm$1.7   & 99.1$\pm$0.9  &  94.9 &    91.0$\pm$1.7   & 92.8$\pm$2.8 & 87.4$\pm$4.9  & \textbf{91.9}  & 92.0$\pm$1.6     & 36.5$\pm$8.6       & \textbf{64.3}  \\ \bottomrule
\end{tabular}
\label{tab:results_op_yamnet}
\end{table*}

\section{Results}\label{sec:results}

The aim of the experiments is to test the capability of the \highl{baseline systems} to correctly classify the examples corresponding to the set of target pattern sounds (KK classes) while successfully rejecting any sound pertaining to an unwanted class, regardless of whether it belongs to a KU class or a UU class. Therefore, the final accuracy of each experiment must take into account not only the correct classification of KK sounds, but also the rejection of any unwanted sounds. 

Following the criteria of Task~1C of DCASE-2019 \cite{b32}, the $ACC_{w}$ measure is used, which is calculated using the following weighted average:
\begin{subequations}\label{eq:accuracy}
    \begin{align}
    \begin{split}
    O^* = 0&~\textrm{(without UU)}:\\
    &ACC_{w} = wACC_{KK} + (1-w)ACC_{KU},\label{eq:accuracy_ku}
    \end{split}
    \end{align}
    \begin{align}
    \begin{split}
    O^* \neq 0&~ \textrm{(with KU and UU)}:\\
    &ACC_{w} = wACC_{KK} + (1-w)ACC_{KUU},\label{eq:accuracy_kuu}
    \end{split}
    \end{align}
    \begin{align}
    \begin{split}
    O^* \neq 0&~ \textrm{(with only UU)}:\\   
    &ACC_{w} = wACC_{KK} + (1-w)ACC_{UU},\label{eq:accuracy_uu}
    \end{split}
    \end{align}
\end{subequations}
where $w$ is an arbitrary weight factor that allows to balance the importance of the accuracy relative to target and unwanted classes. 

\highl{
In the above equations, $ACC_{KK}$ is the multiclass accuracy over 
test examples exclusively from target (KK) classes given as}

\begin{equation}\label{eq:acc}
    \frac{\displaystyle \sum_{k}TP_k}{\displaystyle
    \sum_{k}(TP_k+FN_k)}
\end{equation}

\noindent
\highl{
where $k$ ranges over KK classes and $TP_k$ and $FN_k$ are the number of true positives and false negatives corresponding to class $k$.}
%
Thus, $ACC_{KK} = 1$ when all the test examples belonging to the KK classes are correctly classified. 

\highl{
Correspondingly, $ACC_{KU}$ and $ACC_{UU}$ denote the same accuracy when considering test data either from KU or UU classes and considering two output labels only: pattern and unwanted. They are both obtained using the expression in Equation~\ref{eq:acc} with $k$ ranging over these two class labels on test sets from KU and UU classes, respectively.
}
%

Finally, when the openness is such that there are both KU and UU classes, then the rejection capability is measured by the $ACC_{KUU}$, which is the mean of $ACC_{KU}$ and  $ACC_{UU}$. In the present work $w$ has been given a fixed value of $w=0.5$.  Note that the formulas in Eq.~(\ref{eq:accuracy}) take into account accuracies of all the categories, KK, KU and UU. Therefore, it is a convenient way of analyzing the trade-off between correct prediction and rejection.


Results are presented following $k$-fold cross-validation as indicated in Sec.~\ref{sec:dataset}. For statistical reasons, the folds were run 5 times each. 
\highl{All the tables
show the mean accuracy and standard deviation across all runs and folds. Best performance between the two proposed baseline systems is highlighted using bold typeface.}

\begin{table*}{}
\caption{Baseline L$^3$-net system average accuracies (\%) and corresponding standard deviations (not shown for $ACC_{w}$) for the second batch of experiments using trios (only 3 KK classes).}
\centering
\small
\setlength\tabcolsep{0.5pt}
\begin{tabular}{cccc >{}cccc >{}ccc >{}c}
\toprule
&   & \multicolumn{9}{c}{Openness coefficient}  \\ \cmidrule(lr){3-12}
Trio & Shots   & \multicolumn{3}{c}{$O^* = 0$}    & \multicolumn{4}{c}{$O^* = 0.13$} & \multicolumn{3}{c}{$O^* = 0.39$} \\
\midrule
\multicolumn{1}{l}{}   & \multicolumn{1}{l}{}  & \hspace*{3mm}$ACC_{KK}$\hspace*{3mm}  & $ACC_{KU}$  & $ACC_w$ & $ACC_{KK}$  & \hspace*{2mm}$ACC_{KUU}$ \hspace*{2mm} & $ACC_{UU}$ & $ACC_w$ & \hspace*{3mm}$ACC_{KK}$\hspace*{3mm}    & $ACC_{UU}$   & $ACC_w$  \\
\cmidrule(lr){3-5} \cmidrule(lr){6-9} \cmidrule(lr){10-12}
  \multirow{2}{*}{0} & 1  & 65.1$\pm$16.1 & 99.4$\pm$1.1 & 82.3 &  85.9$\pm$13.4 & 97.7$\pm$4.6 & 98.4$\pm$4.1 & \textbf{91.8} & 100$\pm$0 & 18.6$\pm$8.9  &   \textbf{59.3}  \\ 
  & 2 & 80.2$\pm$15.0 & 99.6$\pm$0.5 & 89.9  & 89.2$\pm$12.5 & 99.6$\pm$0.5 & 99.8$\pm$0.6 & 94.4 & 100$\pm$0 & 17.0$\pm$5.9 & \textbf{58.5} \\ 
  (1, 9, 17) & 4 & 90.1$\pm$14.5 & 99.7$\pm$0.4 & 94.9 & 97.5$\pm$8.1 & 99.7$\pm$0.4 & 99.9$\pm$0.4 & \textbf{98.6} & 100$\pm$0 & 16.9$\pm$3.3 & \textbf{58.5} \\ 
 \cmidrule(lr){2-12}
 \multirow{2}{*}{1} & 1 & 68.9$\pm$12.9 & 99.9$\pm$0.2 & 84.4 & 88.8$\pm$13.1 & 98.3$\pm$2.8 & 96.8$\pm$5.6 & \textbf{93.5} & 100$\pm$0 & 3.9$\pm$3.1 & 52.0 \\ 
     & 2 & 84.7$\pm$16.5 & 99.9$\pm$0.3 & 92.3 & 89.0$\pm$14.5 & 98.7$\pm$2.4 & 97.6$\pm$4.7 & \textbf{93.8} & 100$\pm$0 & 3.6$\pm$2.6 & 51.8 \\ 
     (10, 12, 19)& 4 & 88.0$\pm$15.6 & 99.9$\pm$0.4 & 93.9 & 96.2$\pm$9.6 & 96.7$\pm$3.1 & 93.8$\pm$5.8 & \textbf{96.5} & 100$\pm$0 & 3.8$\pm$3.5 & 51.9 \\ 
     \cmidrule(lr){2-12}
      \multirow{2}{*}{2} & 1 & 55.5$\pm$18.6 & 99.9$\pm$1.0 & 77.7 & 78.4$\pm$13.4 & 99.8$\pm$0.9 & 99.7$\pm$1.7 & 89.1 & 98.6$\pm$2.4  & 14.8$\pm$12.1 & \textbf{56.7} \\ 
   & 2 & 76.1$\pm$14.7 & 99.9$\pm$0.1 & 88.0 & 82.6$\pm$13.9 & 99.8$\pm$0.5 & 99.7$\pm$0.6 & 91.2 & 99.5$\pm$1.2 & 15.7$\pm$11.9 & \textbf{57.6} \\ 
    (2, 14, 22) & 4 & 83.1$\pm$20.7 & 99.9$\pm$0.1 & 91.5 & 91.9$\pm$12.3 & 99.4$\pm$0.9 & 99.0$\pm$1.5 & 95.6 & 99.9$\pm$0.4 & 11.5$\pm$8.2 &  \textbf{55.7} \\ 
     \cmidrule(lr){2-12}
      \multirow{2}{*}{3} & 1 & 53.0$\pm$12.1 & 99.9$\pm$0.4 & 76.5 & 72.3$\pm$13.4 & 96.2$\pm$4.2 & 92.7$\pm$8.2 &  \textbf{84.3} & 99.7$\pm$0.7 & 24.9$\pm$8.2 & \textbf{62.3} \\ 
     & 2 & 64.6$\pm$16.1 & 99.9$\pm$0.3 & 82.2 & 78.4$\pm$13.7 & 95.7$\pm$4.6 & 91.6$\pm$8.7 & \textbf{87.2} & 99.8$\pm$0.5 & 23.3$\pm$6.1 & \textbf{61.6} \\ 
     (3, 6, 13)& 4 & 77.4$\pm$19.0 & 99.8$\pm$0.9 & 88.6 & 90.3$\pm$11.4 & 92.0$\pm$3.2 & 84.8$\pm$6.0 & 91.1 & 99.8$\pm$0.4 & 24.5$\pm$6.0 & \textbf{62.2} \\ 
     \cmidrule(lr){2-12}
      \multirow{2}{*}{4} & 1 & 71.7$\pm$15.2 & 100$\pm$0 & \textbf{85.8} & 88.5$\pm$10.1 & 99.3$\pm$1.3 & 98.6$\pm$2.5 & \textbf{93.9} & 99.8$\pm$0.8 & 2.4$\pm$2.4 & \textbf{51.1} \\ 
     & 2 & 86.8$\pm$14.5 & 100$\pm$0 & \textbf{93.4} & 93.2$\pm$9.2 & 99.4$\pm$1.1 & 98.8$\pm$2.2 & \textbf{96.3} & 100$\pm$0.2 & 1.7$\pm$1.7 & 50.8 \\ 
     (4, 5, 16) & 4 & 88.1$\pm$18.6 & 99.9$\pm$0.6 & 94.0 & 97.0$\pm$9.1 & 99.0$\pm$1.2 & 98.1$\pm$2.2 & \textbf{98.0} & 100$\pm$0 & 1.7$\pm$1.2 & 50.9 \\ 
     \cmidrule(lr){2-12}
       \multirow{2}{*}{5} & 1 & 76.5$\pm$15.2 & 99.9$\pm$0.2 & 88.2 & 87.9$\pm$11.8 & 99.1$\pm$1.2 & 98.5$\pm$2.2 & 93.5 & 97.3$\pm$5.1 & 42.1$\pm$20.1 & \textbf{69.7} \\ 
      & 2 & 85.1$\pm$15.4 & 99.9$\pm$0.1 & 92.5 & 93.4$\pm$7.7 & 98.8$\pm$1.2 & 97.8$\pm$2.3 & 96.1 & 99.1$\pm$2.6 & 39.1$\pm$19.8 & \textbf{69.1} \\ 
      (18, 21, 23) & 4 & 89.3$\pm$16.4 & 100$\pm$0.1 & 94.6 & 97.2$\pm$8.1 & 98.3$\pm$1.2 & 96.8$\pm$2.1 & 97.7 & 99.9$\pm$0.3 & 34.3$\pm$20.2 & \textbf{67.1} \\ 
     \cmidrule(lr){2-12}
      \multirow{2}{*}{6} & 1 & 87.0$\pm$13.5 & 99.7$\pm$0.5 & 93.4 & 96.0$\pm$7.8 & 99.3$\pm$0.8 & 99.4$\pm$0.6 & \textbf{97.6} & 100$\pm$0 & 30.9$\pm$11.6 & \textbf{65.5} \\ 
  & 2 & 87.6$\pm$16.0 & 99.6$\pm$0.6 & 93.6 & 95.8$\pm$9.1 & 99.4$\pm$0.7 & 99.2$\pm$1.0 & 97.6 & 100$\pm$0 & 28.2$\pm$9.5 & \textbf{64.1} \\ 
      (8, 11, 24)& 4 & 89.9$\pm$14.5 & 99.7$\pm$0.5 & 94.8 & 96.8$\pm$9.2 & 99.2$\pm$0.8 & 98.9$\pm$1.0 & 98.0 & 100$\pm$0 & 27.7$\pm$8.0 & \textbf{63.9} \\ 
     \cmidrule(lr){2-12}
      \multirow{2}{*}{7} & 1 & 66.4$\pm$15.7 & 99.6$\pm$0.6 & 83.0 & 87.0$\pm$11.4 & 97.6$\pm$2.9 & 96.8$\pm$5.4 & \textbf{92.3} & 99.2$\pm$1.9  & 23.7$\pm$8.0 & \textbf{61.5} \\ 
      & 2 & 82.1$\pm$13.7 & 99.5$\pm$0.7 & 90.8 & 90.0$\pm$9.8 & 98.6$\pm$1.7 & 98.4$\pm$3.0 & 94.3 & 99.8$\pm$0.6 & 24.0$\pm$6.7 & 61.9 \\ 
      (7, 15, 20) & 4 & 83.7$\pm$15.3 & 99.5$\pm$0.9 & 91.6 & 94.4$\pm$10.1 & 98.5$\pm$1.5 & 98.1$\pm$2.7 & 96.5 & 100$\pm$0.2 & 24.2$\pm$5.3 & 62.1 \\ \bottomrule
\end{tabular}
\label{tab:L$^3$-net_3}
\end{table*}

\begin{table*}{}
\caption{Baseline YAMNet system average accuracies (\%) and corresponding standard deviations (not shown for $ACC_{w}$) for the second batch of experiments using trios (only 3 KK classes).}
\centering
\small
\setlength\tabcolsep{0.5pt}
\begin{tabular}{cccc >{}cccc >{}ccc >{}c}
\toprule
&   & \multicolumn{9}{c}{Openness coefficient}  \\ \cmidrule(lr){3-12}
Trio & Shots   & \multicolumn{3}{c}{$O^* = 0$}    & \multicolumn{4}{c}{$O^* = 0.13$} & \multicolumn{3}{c}{$O^* = 0.39$} \\
\midrule
\multicolumn{1}{l}{}   & \multicolumn{1}{l}{}  & \hspace*{3mm}$ACC_{KK}$\hspace*{3mm}  & $ACC_{KU}$  & $ACC_w$ & $ACC_{KK}$  & \hspace*{2mm}$ACC_{KUU}$ \hspace*{2mm} & $ACC_{UU}$ & $ACC_w$ & \hspace*{3mm}$ACC_{KK}$\hspace*{3mm}    & $ACC_{UU}$   & $ACC_w$  \\
\cmidrule(lr){3-5} \cmidrule(lr){6-9} \cmidrule(lr){10-12}
  \multirow{2}{*}{0} & 1  & 83.8$\pm$9.4 & 97.3$\pm$3.3 & \textbf{90.6}  &  87.0$\pm$8.7 & 92.3$\pm$4.5 & 90.6$\pm$5.6 & 89.6 & 94.0$\pm$7.1 & 17.3$\pm$13.0  & 55.6   \\ 
  & 2 & 93.6$\pm$4.4 & 99.4$\pm$0.8 & \textbf{96.5}  & 94.2$\pm$4.9 & 94.9$\pm$3.9 & 92.5$\pm$5.3 & \textbf{94.5} & 97.4$\pm$3.1 & 16.1$\pm$11.0 & 56.7 \\ 
  (1, 9, 17) & 4 & 97.8$\pm$3.0 & 99.8$\pm$0.4 & \textbf{98.8} & 97.7$\pm$3.4 & 96.5$\pm$2.3 & 94.1$\pm$3.5 & 97.1 & 98.6$\pm$2.8 & 17.0$\pm$17.0 & 57.8 \\ 
 \cmidrule(lr){2-12}
 \multirow{2}{*}{1} & 1 & 83.9$\pm$5.7 & 96.5$\pm$3.8 & \textbf{90.2} & 88.2$\pm$5.9 & 91.7$\pm$4.1 & 89.5$\pm$5.7 & 90.0 & 96.0$\pm$2.4 & 26.2$\pm$14.9 & \textbf{61.1} \\ 
     & 2 & 92.8$\pm$4.8 & 99.4$\pm$1.1 & \textbf{96.1} & 92.6$\pm$5.9 & 91.6$\pm$4.7 & 87.8$\pm$6.8 & 92.1 & 97.2$\pm$2.5 & 25.4$\pm$16.6 & \textbf{61.3} \\ 
     (10, 12, 19)& 4 & 96.5$\pm$2.7 & 99.8$\pm$0.3 & \textbf{98.2} & 96.4$\pm$2.3 & 95.1$\pm$3.3 & 91.2$\pm$5.8 & 95.7 & 98.0$\pm$2.2 & 21.6$\pm$14.4 & \textbf{59.8} \\ 
     \cmidrule(lr){2-12}
      \multirow{2}{*}{2} & 1 & 96.9$\pm$3.7 & 99.9$\pm$0.1 & \textbf{98.4 }& 97.7$\pm$4.8 & 97.7$\pm$3.0 & 95.9$\pm$3.7 & \textbf{97.7} & 98.7$\pm$5.5  & 11.0$\pm$7.2 & 54.8 \\ 
   & 2 & 98.4$\pm$1.1 & 100$\pm$0 & \textbf{99.2} & 99.2$\pm$0.8 & 97.6$\pm$1.7 & 95.4$\pm$3.2 & \textbf{98.4} & 100$\pm$0 & 8.3$\pm$7.2 & 54.2 \\ 
    (2, 14, 22) & 4 & 98.9$\pm$1.1 & 100$\pm$0 & \textbf{99.5} & 99.4$\pm$0.8 & 97.1$\pm$1.1 & 94.4$\pm$2.2 & \textbf{98.2} & 100$\pm$0 & 4.9$\pm$5.8 & 52.5 \\ 
     \cmidrule(lr){2-12}
      \multirow{2}{*}{3} & 1 & 58.9$\pm$7.9 & 95.5$\pm$3.4 & \textbf{77.2} & 63.1$\pm$8.2 & 89.6$\pm$3.7 & 86.4$\pm$4.7 & 76.3  & 68.9$\pm$7.3 & 5.2$\pm$6.1 & 37.0 \\ 
     & 2 & 70.8$\pm$7.0 & 98.3$\pm$1.5 & \textbf{84.5} & 73.6$\pm$6.1 & 91.7$\pm$3.4 & 88.3$\pm$4.5 & 82.7 & 79.0$\pm$6.4 & 7.5$\pm$8.2 & 43.3 \\ 
     (3, 6, 13)& 4 & 85.9$\pm$5.2 & 99.6$\pm$0.6 & \textbf{92.7} & 86.8$\pm$4.7 & 95.9$\pm$2.7 & 93.1$\pm$4.2 & \textbf{91.4} & 94.1$\pm$4.3 & 8.8$\pm$5.2 & 51.5 \\ 
     \cmidrule(lr){2-12}
      \multirow{2}{*}{4} & 1 & 71.1$\pm$8.2 & 97.6$\pm$3.0 & 83.7 & 75.1$\pm$8.1 & 92.4$\pm$4.4 & 90.0$\pm$5.3 & 83.7 & 82.1$\pm$8.2 & 12.4$\pm$12.5 & 47.3 \\ 
     & 2 & 85.7$\pm$6.1 & 99.1$\pm$1.1 & 92.4 & 88.8$\pm$5.9 & 92.5$\pm$3.6 & 88.2$\pm$5.9 & 90.7 & 93.8$\pm$6.2 & 10.6$\pm$9.0 & \textbf{52.2} \\ 
     (4, 5, 16) & 4 & 92.1$\pm$5.1 & 99.9$\pm$0.2 & \textbf{96.0} & 93.4$\pm$4.9 & 93.4$\pm$3.9 & 88.6$\pm$6.8 & 93.4 & 97.6$\pm$3.3 & 9.9$\pm$6.2 & \textbf{53.8} \\ 
     \cmidrule(lr){2-12}
       \multirow{2}{*}{5} & 1 & 98.6$\pm$5.1 & 99.7$\pm$1.0 & \textbf{99.2} & 99.7$\pm$1.7 & 99.6$\pm$1.9 & 99.6$\pm$1.5 & \textbf{99.6} & 100$\pm$0 & 24.3$\pm$13.8 & 62.2 \\ 
      & 2 & 99.6$\pm$2.1 & 100$\pm$0 & \textbf{99.8} & 99.9$\pm$0.5 & 99.9$\pm$0.2 & 99.9$\pm$0.2 & \textbf{99.9} & 100$\pm$0 & 20.9$\pm$12.7 & 60.4 \\ 
      (18, 21, 23) & 4 & 100$\pm$0 & 100$\pm$0 & \textbf{100} & 100$\pm$0 & 100$\pm$0 & 100$\pm$0 & \textbf{100} & 100$\pm$0 & 21.3$\pm$15.4 & 60.8 \\ 
     \cmidrule(lr){2-12}
      \multirow{2}{*}{6} & 1 & 94.2$\pm$7.8 & 99.6$\pm$1.4 & \textbf{97.0} & 94.6$\pm$6.1 & 98.1$\pm$2.6 & 96.8$\pm$3.5 & 96.4 & 96.4$\pm$3.2 & 14.6$\pm$7.1 & 55.5  \\ 
  & 2 & 98.0$\pm$4.2 & $100\pm$0 & \textbf{99.0} & 98.1$\pm$3.1 & 98.8$\pm$0.9 & 97.7$\pm$1.9 & \textbf{98.4} & 97.5$\pm$2.9 & 12.4$\pm$5.9 & 55.0 \\ 
      (8, 11, 24)& 4 & 99.4$\pm$1.4 & 100$\pm$0 & \textbf{99.7} & 99.4$\pm$1.5 & 98.8$\pm$0.8 & 97.8$\pm$1.5 & \textbf{99.1} & 98.7$\pm$2.7 & 11.0$\pm$4.7 & 54.8 \\ 
     \cmidrule(lr){2-12}
      \multirow{2}{*}{7} & 1 & 86.1$\pm$9.2 & 98.7$\pm$2.3 & \textbf{92.4} & 86.3$\pm$9.5 & 96.4$\pm$3.9 & 96.7$\pm$3.7 & 91.4 & 88.8$\pm$9.3  & 25.8$\pm$13.1 & 57.3 \\ 
      & 2 & 93.2$\pm$4.5 & 99.6$\pm$0.7 & \textbf{96.4} & 93.4$\pm$4.2 & 98.0$\pm$2.4 & 98.3$\pm$1.9 & \textbf{95.7} & 94.6$\pm$3.4 & 31.8$\pm$11.4 & \textbf{63.2} \\ 
      (7, 15, 20) & 4 & 96.0$\pm$2.4 & 99.7$\pm$0.6 & \textbf{97.8} & 96.0$\pm$2.3 & 99.3$\pm$0.8 & 99.6$\pm$0.4 & \textbf{97.6} & 96.2$\pm$2.5 & 35.9$\pm$9.7 & \textbf{66.1} \\ \bottomrule
\end{tabular}
\label{tab:yamnet_3}
\end{table*}

\subsection{Large number of target classes}

\highl{The results obtained by the two baseline systems for the first batch of experiments are shown in Tables ~\ref{tab:results_op} and ~\ref{tab:results_op_yamnet}. In this first batch, the KK set comprises the 24 pattern sound classes. As indicated in Table~\ref{tab:classes_openness}, three values of openness are considered: $O^*\in \{0,\, 0.04,\, 0.09\}$. As expected, the results confirm, in general, the difficulties encountered in FSL and OSR conditions. On the one hand, the lack of a large number of training examples affects considerably the classification performance, as evidenced, for example, by the low $ACC_w$ values achieved by the L$^3$-net system when only one shot is used for training. On the other hand, as the openness value increases, the accuracy for KK classes remains similar whereas the accuracy of KU-UU classes decreases.} 

\highl{Low values in $ACC_{UU}$ and/or $ACC_{KU}$ indicate that the system is misclassifying unwanted events as \emph{pattern sounds}, meaning that false positives are observed in the KK categories. As expected, the problems arising from UU classes are more evident under higher openness conditions. By letting the system learn from a set of unwanted sounds, the rejection capabilities are considerably increased. This is evidenced by the higher values in $ACC_{UU}$ for $O^{*}=0.04$ with respect to the ones for $O^{*}=0.09$, independently of the baseline system used. Note, however, that the use of unwanted sounds for training the classifier may also have an impact in the accuracy achieved for the target pattern sounds. As shown in both tables, at $O^{*}=0$, the accuracy for the KK classes is worse than for higher openness. This is because the use of $KU$ classes to train the system makes the underlying classification boundaries more restrictive, and the system is more prone to miss target instances.} 

\highl{In general terms, YAMNet shows a greater weighted accuracy regarding known and unknown situations when $O^*\in \{0.04,\, 0.09\}$. Thus, YAMNet could be understood as a more discriminative extractor when unknown situations are present. However, the most significant phenomenon can be seen when $O^*=0$ and the number of shots is equal to 1. A huge improvement in $ACC_{KK}$ is observed with respect to L$^{3}$-net, leading to a better trade-off in $ACC_{w}$. The improvement of this feature extractor is nearly of 25 percentage points (see Table~\ref{tab:results_op}). The difference between 1 shot and 2 shots with $O^*=0$ using L$^3$-net is more than 30 percentage points, while for YAMNet is only of 8 percentage points. Therefore, YAMNet seems to be a more robust solution.}

\subsection{Small number of target classes}

\highl{Tables~\ref{tab:L$^3$-net_3} and \ref{tab:yamnet_3} show the results for the second batch of experiments that consider only KK sets comprised of 3 pattern sound classes, considering 8 different and disjoint trios. As indicated in Table~\ref{tab:classes_openness}, the three values of openness in this case are:  $O^*\in \{0,\, 0.13,\, 0.39\}$. Again, the general tendency is confirmed, where a lower number of shots or a higher openness level leads always to a decrease in performance. However, in this case, it can be observed that the particularities of the target classes can be also an important factor affecting the overall performance of the system. For example, with $O^* = 0.39$, very low values for $ACC_{UU}$ are obtained in Table~\ref{tab:L$^3$-net_3} for trios 1 and 4, considerably worse than for other trios in the dataset. The internal L$^3$-net representations of such target classes may probably lead to classification boundaries that are not discriminatory enough to reject successfully the unwanted sounds. Interestingly, the specific internal representations are also of high importance, as the same trios are not the ones with lowest performance in YAMNet (see Table~\ref{tab:yamnet_3}).}

\highl{In any case, the differences between the two baseline systems are much more evident in this second batch of experiments than in the previous one. While  $O^*=0$ was the case that most favored the L$^3$-net baseline when $|KK|=24$, with trios YAMNet seems to offer better performance for the same value of openness. The tendency is also reversed for the highest level of openness ($O^*=0.39$), as the L$^3$-net embeddings show now the best performance for most trios. Finally, note that the trio-wise results are quite balanced for $O^*=0.13$, as both systems are similarly competitive. However, the winning system is again quite dependent on the actual trio.}

\section{Conclusion and future work}\label{sec:conclusion}

\par Few-shot learning (FSL) is a research area with increasing interest in the audio domain. However, the lack of public FSL audio datasets makes it necessary to manipulate other existing databases with the aim of adapting them properly to FSL research. Moreover, open-set recognition (OSR) can be an additional problem in practical FSL scenarios, where the models are likely to be tested with instances from unseen classes during training. This work presented a carefully designed audio dataset for FSL and OSR research, where target sounds are instances of classes corresponding to different audio patterns (fire alarms, doorbells, etc.). The dataset considers a domestic scenario where such audio pattern classes correspond to intentional sounds to be accurately detected in the presence of other unwanted sounds (coughs, door slams, etc.). Each class comes with different samples for FSL training, validation and testing, under different openness conditions. To facilitate the use of this dataset and promote algorithm development, we also provide results with a baseline system using transfer learning from \highl{pre-trained state-of-the-art convolutional neural networks}. The results show that important trade-offs exist when both FSL and OSR conditions are considered, evidencing the need for novel learning architectures aimed at facing both types of problems. \highl{Future updates of this dataset will include more challenging acoustic conditions, such as different levels of noise, reverberation and overlapped events}.

\section*{Acknowledgment}

This project has received funding from the European Union’s Horizon 2020 research and innovation program under grant agreement No 779158. The participation of Javier Naranjo-Alcazar and Dr. Pedro Zuccarello in this work is partially supported by Torres Quevedo fellowships DIN2018-009982 and PTQ-17-09106 respectively from the Spanish Ministry of Science, Innovation and Universities. The participation of Dr. Cobos  and Dr. Ferri is supported by ERDF and the Spanish Ministry of Science,
Innovation and Universities under Grant RTI2018-097045-B-C21, as well as grants AICO/2020/154 and AEST/2020/012 from Generalitat Valenciana.





\bibliographystyle{unsrt}
\bibliography{references}

\end{document}